\magnification = \magstep 1
\def\ctr#1{\centerline{#1}}
\def\resumen#1{{\narrower #1}}
\def\titulo#1{\vskip 5mm #1 \vskip 2mm}
\def\ecuacion#1#2#3{$$ #1 \, #2 \eqno (#3) $$}
\def\ecuarreglo#1#2#3{$$\eqalignno{#1 \, #2 & (#3) \cr}$$}
\def\lag{{\cal L}}
\def\func{{\cal F}}
\def\diag{\mathop {\rm diag}}
\def\dvec#1{{\buildrel \leftrightarrow \over #1}}

{\bf \ctr{A Charged Inflaton leaves Behind a Fractal}
\ctr{Primeval Structure of Electromagnetic Fields}}
\vskip 5mm
\ctr{M. Chaves* and G. Padilla$^{\rm \dag}$}
{\it \ctr{Escuela de F\'{\i}sica, Universidad de Costa Rica}
\ctr{Ciudad Universitaria Rodrigo Facio, San Jos\'e, Costa Rica}}
\vskip 5mm
\resumen{
The inflaton field is assumed to possess electric charge. The effect of
the charge upon inflation is studied using a parameter $\omega_0$ that acts as
a measure of the quantity of inflaton present. It is shown that at the end of 
inflation the charged inflaton should produce a fractal electromagnetic
structure, which could act as a seed for the development of macroscopic 
galactic structures. Another consequence would be the presence nowadays 
of residual electromagnetic fields that would encompass galaxies, clusters, 
and larger structures.
\vskip 5mm
PACS numbers: 98.65.Dx, 98.80.Cq, 11.10.Ef, 41.20.-q
}

\titulo{1. Introduction}

The Big-Bang Model of cosmology has such a strong experimental support that 
the odds are that it is fundamentally correct. But if the expansion factor 
of the universe has always been a power of the time, that is, if
\ecuacion{a(t) \propto t^n}{}{1}
always, then certain theoretical difficulties arise, some of which (perhaps 
the most important) are:
\item{A.} An unexplained large entropy density at the beginning of the universe, or, 
what is logically equivalent, the so-called flatness problem, which is the 
observation that there is no explanation for the fact the density parameter 
$\Omega \equiv \rho / \rho_c$ is {\it so close to one} at early times.
\item{B.} The homogeneity of our present universe, which seems to contradict the 
fact that it is made up of many patches that have {\it always} been causally 
disconnected.
\item{C.} The absence in our universe of monopoles and other relics that should 
have been created during early times.

\noindent
While these conditions do not contradict the theoretical picture of the 
Big-Bang Model, they are not explained by it, and point to its incompleteness.

The inflation paradigm [1] postulates a scalar field $\varphi$, called the inflaton, 
which possesses a self-interacting potential $V(\varphi)$. Under suitable circumstances, 
the pressure $p$ produced by this field is negative, so that the work done by 
the expanding universe is negative and it exactly compensates the energy 
necessary to maintain a constant energy density while the volume continuously 
grows. According to the inflationary paradigm, at the beginning of this 
universe, or shortly after, the inflaton's energy dominated over other 
types and as a result there was an exponential expansion
\ecuacion{a(t) = {\rm ct.} \times e^{\alpha t}}{}{2}
through which the universe becomes many times larger. This growth explains 
the previously mentioned theoretical peculiarities as follows: The 
inflationary expansion, which is caused by a cosmic potential energy density, 
drives the density parameter $\Omega$ to exactly the value one (as can be proven 
using the equations of general relativity). This same energy density has 
then to be converted into elementary particles in a dissipative process that 
is very generous in its entropy production. The expansion is so large that 
the entire observable universe comes from only one small patch in the early 
universe, and monopoles and other relics are spread over such a vast volume 
that they become unobservable from a practical point of view. So all the 
difficulties we listed are resolved.

At inflation's end all that is left is the vacuum and the quantum fluctuations 
of the inflaton field. All the cosmic structure seen today would have to 
originate in these fluctuations. In the last few years redshift surveys of 
large numbers of galaxies [2] have shown the universe to possess hitherto 
unsuspected macrostructures. The galaxies are distributed along the walls of 
huge voids as large as 100 Megaparsecs forming a honeycomb structure. It 
could be possible that even larger structures can become apparent with a 
survey encompassing a larger volume. At present there is much interest in 
this question, and two major redshift surveys are underway. [3] Just how 
consistent the known macrostructures are with the usual inflation scenario, 
where the only possible source for structure are the inflaton's quantum 
fluctuations, is not clear at present. But if even larger structures are 
found, it will be impossible to accommodate them within this scenario.

In this paper we are going to assume the inflaton to be electrically charged 
(positively, for simplicity) and study how its charge is going to affect the 
inflationary process. Besides being an alternative that should be worked out, 
there is a logical reason to undergo such a study. It seems that scalar 
bosons can have a dynamic character, that is, they can be gauge bosons. [4]
In this case, in the context of a generalized grand unified theory, they 
would couple to the gauge vector bosons and thus to the photon. Since the 
photon is massless, the inflaton is going to produce EM fields far more than 
any other kind of field.

The charge of the inflaton requires it to be a complex field. Incidentally, 
we shall assume a curved spacetime with a metric $g_{\mu \nu} = \diag (1, -a^2, -a^2, -a^2)$
and a time coordinate $x^0$. 
We shall frequently use the 4-current $j_\mu=i\psi^*\dvec{\partial}_\mu\psi=(j_0,{\bf j})$, 
where $\dvec{\partial}_\mu \equiv \partial_\mu - {\buildrel \leftarrow \over \partial}_\mu$. Complex inflatons have 
been treated before in the literature, but with different interests in 
mind. [5] In our treatment we include in the Lagrangian of the boson 
quantum theory a term $\omega_0 j_0$, that classifies homogenous states of complex bosons. 
Such states are exact solutions of $T=0$ quantum field theories and are 
metastable under certain conditions. [6] The $\omega_0$ acts as a chemical potential 
in the sense that it is a measure of the amount of charge there is. 
This term is important in the understanding of the physical picture since, 
as we shall see, it forces the creation of electromagnetic (EM) fields as 
the universe expands. What this means physically is that the charged 
inflaton is forced to convert part of its energy into EM fields as the 
universe expands during inflation. Some of these EM fields have sizes 
comparable to the size of the patch that eventually results in our universe. 
Thus at inflation's end there would exist in the universe a fractal array 
of EM fields that can act as seed for macrostructures.

\titulo{2. Inflation with a complex inflaton}

A typical Lagrangian for a complex field in flat spacetime is
\ecuacion{\lag = |\partial_0 \varphi|^2 - |\nabla \varphi|^2 - m^2|\varphi |^2 -\lambda |\varphi|^4}{.}{3}
The mass term has the usual sign, resulting in a potential with a minimum 
at $\varphi=0$. An imaginary mass, on the contrary, would result in a minimum at a 
nonzero value for $\varphi$. Recently the metastability of certain $T=0$ solutions of 
the system given by (3) has been proven under fairly common conditions. [6] 
The metastability is brought into evidence by means of a field theory 
contact transformation, and the solutions are classified by a frequency
$\omega_0$ that acts as a chemical potential (although there is no ensemble 
involved here) in the sense that it measures how much of the field there is. 
The last term in (3) is a self-interacting term that we will not include in 
our analysis.

We are interested in an electrically charged inflaton field $\varphi$ that interacts 
with the EM potential $A_\mu$, and so we use the quantum field theory Lagrangian:
\ecuacion{\lag = |(i \partial_\mu - eA_\mu)\varphi|^2 - m^2|\varphi |^2 - {1 \over 4}F_{\mu \nu}F^{\mu \nu}}{.}{4}
This Lagrangian possess the local $U(1)$ gauge invariance appropriate to 
electromagnetism. As an initial condition we assume that there is a uniform 
inflaton charge spread throughout the whole universe. We would like to 
measure the inflaton's charge density in some convenient way. A reliable, 
if somewhat abstract way of doing this, is to perform the contact 
transformation (more specifically, the point transformation) generated by 
the functional
\ecuacion{\func = e^{i \omega_0 t} \Pi \varphi + e^{-i \omega_0 t} \varphi^* \Pi^*}{}{5}
between the fields $(\varphi, \pi)$ and $(\psi, \Pi)$, where $\pi=\partial \lag/\partial \dot{\varphi}$
and $\Pi=\partial \lag'/\partial \dot{\psi}$. [6] The resulting transformed 
Lagrangian, written with the metric in an expanded form, is:
\ecuarreglo{\lag' = &\ |\dot{\psi}|^2 - a^{-2}|\nabla \psi|^2 + [(\omega_0 - eA_0)^2 - m^2 -a^{-2}e^2{\bf A \cdot A}]|\psi|^2 \cr%
 &+ (\omega_0 - e A_0) j_0 + a^{-2} e {\bf A \cdot j} - {1 \over 4} F_{\mu \nu} F^{\mu \nu}}{.}{6}
The parameter $\omega_0$ serves then as a measure of the amount of charge present.

We need the Euler-Lagrange equations for the field $\psi$, but one may wish first 
to add the term $\Delta \lag'=-\lambda(j_0)^2$ to this Lagrangian. This term
represents the repulsive electric self-interaction of the charge distribution 
$j_0$, and it takes such a simple form only when the distribution is fairly 
uniform. The electric energy density at the origin would be
\ecuacion{\int d^3 x' j_0 ({\bf 0}) {e^2 \over |{\bf x'}|} j_0 ({\bf x'}) \approx \lambda [j_0 ({\bf 0})]^2 }{,}{7}
where $\lambda$ is an effective coupling constant that has a value depending on the 
size and type of universe. Technically speaking, this term arises from the 
quantum field theory Lagrangian as a second-order term in the perturbative 
expansion. The Euler-Lagrange equation of motion is:
\ecuarreglo{0 =& -2i \lambda (\partial_0 j_0) \psi - 4i \lambda j_0 \dot{\psi} + \ddot{\psi} + [3H - 2i(\omega_0 - eA_0)]\dot{\psi} - 2ia^{-2}e{\bf A \cdot}\nabla \psi - a^{-2}\nabla^2 \psi \cr%
 &+[-3Hi(\omega_0 - eA_0) + e^2a^{-2}{\bf A \cdot A} + ie\dot{A}_0 - ia^{-2}e\nabla \cdot {\bf A} - (\omega_0 - e A_0)^2 + m^2]\psi}{,}{8}
where $H=\dot{a}/a$ and the two first terms on the right are the ones due to $\Delta \lag'$. 
Our problem is to solve this equation simultaneously with the equations of 
electromagnetism and of general relativity.

The simplest solution, and the one that has evident physical interest, is 
the one where imaginary terms like $-3Hi(\omega_0-eA_0)\psi$ are basically zero for the short while 
inflation lasts because $\omega_0\approx eA_0$. The inflaton's charged current produces EM fields 
that contain a relatively small part of its energy, but are of theoretical 
interest. The resulting EM structure does not affect very much the dynamics 
of inflation for two reasons: first, is coupled to the inflaton through the 
small fine structure constant, and, second, the induced fields are weakened 
very fast due to the rapid volume growth. The field $\psi$ has a solution that is 
similar to the one Linde employed in his chaotic inflation model. [7] 
Linde's solution requires a large initial value for the inflaton field (of 
the order of Planck's energy) and a steeply descending potential $V(\psi)$, which, 
in our case, is given by the mass term $m^2|\psi|^2$. In it the expansion factor $a$ grows 
exponentially at a fixed rate throughout inflation, as in (2). 
The inflaton {\it begins} the inflationary period {\it decreasing} exponentially at a
rate $\beta$; that is, initially,
\ecuacion{\psi = {\rm ct.} \times e^{-\beta t}}{.}{9}
However, the rate of decrease $\beta$ of the inflaton increases slowly throughout 
the inflationary period until it stops the inflationary process, as we shall 
presently see.

The first two terms on the right in equation (8), that are due to the second 
order contribution $\Delta \lag'$, vanish because both $j_0=i\psi^*\dvec{\partial}_0\psi=0$ and 
$\partial_0 j_0 = i\psi^* \ddot{\psi} - i\ddot{\psi}^* \psi = 0$ are zero 
for an inflaton obeying (9).

The equations of general relativity for a Robertson-Walker universe (with 
zero space curvature and cosmological constant) are
\ecuacion{H^2 = M^{-2}_P \rho\quad{\rm and}\quad\ddot{a}/a = -{1 \over 2} M^{-2}_P (\rho + 3p)}{,}{10}
where $\rho$ is the density, $p$ is the pressure, and $M^{-2}_P\equiv{8 \over 3}\pi G$. 
For the inflationary expansion described by (2), we have that $H^2=\ddot{a}/a=\alpha^2$, 
so we conclude that
\ecuacion{\rho = -p = {\rm ct.}}{,}{11}
which is a necessary and sufficient condition for inflation. Neglecting 
again terms with the factor $(\omega_0-eA_0)$ and the induced EM fields, 
the density is given by $\rho=T_{00}=|\dot{\psi}|^2+m^2|\psi|^2$ and the 
pressure by $p={1 \over 3}T_{kk}=|\dot{\psi}|^2-m^2|\psi|^2$. 
It is evident from these expressions that 
condition (11) is satisfied only if $|\dot{\psi}|^2\ll m^2|\psi|^2$. This is the same as requiring 
$\beta^2 \ll m^2$ initially. Something interesting happens here. Neglecting 
in (8) terms with the factor $(\omega_0-eA_0)$ and also the smaller inhomogeneous, 
anisotropic terms due to induced EM fields, equation (8) can be simplified 
into the form
\ecuacion{0 = \ddot{\psi} + 3H\dot{\psi} + m^2 \psi}{.}{12}
Notice that the same requirement $\beta^2\ll m^2$ that we made to insure that (11) hold 
also assures us that the second time derivative in (12) is initially 
negligible! This, plus the fact that initially $\psi\approx M_P$, are the technical details 
that allow in this case inflation to last long enough to solve many of the 
theoretical problems of the Big-Bang. Now, substituting $\psi$, as given by (9), 
in (12) leads us to the relation $3\alpha\beta=m^2$, from which we conclude that $\beta\ll \alpha$. 
This last inequality tells us that the inflaton field is not going to be 
zero at inflation's end, because its decay rate is far smaller than the 
expansion factor's growth rate.

Particles, monopoles and EM fields are weakened into oblivion due to the 
tremendous inflationary expansion. In traditional inflation only quantum 
fluctuations are left to generate structure; however, if the inflaton has 
an electric charge, it generates EM fields throughout inflation. Of those, 
the ones induced just before the end of the inflationary process remain. 
In other words, the EM fields induced at the end of inflation are not 
weakened by further expansion and can be the seed of a complex structure. 
As we said before, the small decay rate of the inflaton (compared with the 
large expansion rate of the expansion factor) assures us that there is 
going to be inflaton left at inflation's end. From a mathematical point of 
view, the origin of these induced fields that survive inflation can be 
traced to the imaginary terms of equation (8), that have to cancel among 
themselves, and that have $\omega_0$ to give them a permanent scale. 
From a physical point of view, they are the last fields induced by the 
charged inflaton just before inflation's end.

The inflaton's electric charge density left at the end of inflation is furthered 
weakened by the usual Big-Bang expansion, but there should still be an 
asymmetry in the electric charge present nowadays in the cosmos. 
Gradients in space or time of the inflaton produce EM fields that in turn 
can produce other EM or inflatons. (Notice, regarding this point, that 
inflatons can couple directly with two photons, so that it is possible for 
a photon to decay {\it directly} into two inflatons and another photon, as long 
as it has energy $E>2m$.) Therefore, as a result of their interaction with the 
inflaton field, a complicated fractal picture of electric currents and EM 
fields is formed, that ranges from macroscopic EM structures down to 
incoherent photons and particles. It is perhaps possible that this 
thermalization is enough to account for present-day matter, in which case 
no other dissipative mechanism would be necessary. The fractal 
macrostructures resulting from the charged inflaton's expansion call to 
mind previous statistical correlation analysis that did not agree with a 
scale-invariant fluctuation spectrum. [8]

\titulo{3. Concluding remarks}

The Euler-Lagrange equation of Lagrangian (6) for the field $A_0$ is 
\ecuacion{a^{-2}\nabla \cdot (\nabla A_0 + \dot{\bf A}) = - e j_0 + 2e(eA_0 - \omega_0)|\psi|^2}{.}{13}
Notice the strong weakening effect of the $a^{-2}$ on the induced EM fields. 
Thus throughout inflation $\omega\approx eA_0$, but at inflation's end the expansion factor 
will not affect so decisively those fields and $A_0$ will begin to strongly 
depend on time. The question arises as to what direction the field $\bf A$ is 
going to take every time it appears at a different place of what should be 
a fairly isotropic universe. Every such occurrence represents a spontaneous 
breaking of the rotational symmetry. We suspect that quantum fluctuations
$\delta \bf A$ of the vector field must be behind these symmetry breakings.

The scenario we are contemplating is the following. Near the end of 
inflation the vacuum is populated with quantum fluctuations and EM fields. 
Thermalization occurs and the universe begins an expansion that goes as a 
power of time. After 300,000 years go by, the universe becomes transparent 
to electromagnetic radiation and the radiation bath can suddenly propagate 
throughout the whole universe. About $10^{10}$ years later here on earth we study 
the radiation (now redshifted to the microwave) that was emitted precisely 
at this moment. So we are studying the structure impressed on a sphere with 
precisely a $10^{10}$ light-years radius. [9] We see a certain anisotropy of one 
part in 100,000 in the cosmic microwave background radiation, that could be 
due to the inflaton's quantum fluctuations, or perhaps to the currents and 
density gradients induced in the plasma by the EM fields. 

The information we get from the redshift surveys of galaxies has a free 
parameter, the redshift itself. By focusing on a specific redshift, we can 
study the structure of a sphere of a specific radius. By varying the redshift 
we can obtain three-dimensional mappings of the galaxies. The other sphere, 
the microwave one, is much farther away than the ones we study through 
redshifts, and the structure we see, of a two-dimensional nature, has yet to 
be enhanced by the effect of gravity working through billions of years. 
The microwave sphere is a cross section of the possible three-dimensional 
structure that could exist at the time. There seems to be no contradiction, 
at least in principle, between the small anisotropic microwave signal and 
the macroscopic structures seen at optical wavelengths. The data that will 
become available in the next few years will very likely clarify these topics.

One final comment. If our hypothesis is true and the field that drove 
inflation had an electric charge, the EM fields it induced should exist even 
today, although much weakened by the usual Big Bang expansion. They must 
encompass all galaxies, clusters and macrostructures.

\titulo{ACKNOWLEDGEMENT}

We wish to thank Dr. Walter Fern\'andez, head of the Laboratorio de 
Investigaciones Atmosf\'ericas y Planetarias, University of Costa Rica, for 
kindly allowing us to become a rather frequent users of the computers there 
during this past year.

\titulo{REFERENCES}

* Electronic address: mchaves@cariari.ucr.ac.cr

$^{\rm \dag}$ Electronic address: gpadilla@cariari.ucr.ac.cr
\vskip 4mm

\item{[1]} E. W. Kolb, M. S. Turner, {\it The Early Universe}, Addison-Wesley Publishing Company, Redwood City, 1990.
\item{[2]} M. A. Strauss, Nature (1998), in press, special supplementary issue celebrating KPNO's $40^{\rm th}$ aniversary; 
S. A. Schetman et al., Astrophys. J. {\bf 470}, 172 (1996); 
M. S. Vogeley, C. Park, M. J. Geller and J. P Huchra, Astrophys. J. {\bf 420}, 525 (1994); 
M. J. Geller and J. P. Huchra, Science {\bf 246}, 897 (1989); 
V. De Lapparent, M. J. Geller and J. P. Huchra, Astropys. J. {\bf 332}, 44 (1988); 
V. De Lapparent, M. J. Geller and J. P. Huchra, Astrophys. J. {\bf 302}, L1 (1986).
\item{[3]} M. Colles, Phil. Trans. R. Soc. Lond., in press, 1998; 
J. R. Knapp, Sky \& Tel. {\bf 94}, 40 (1997); 
J. E. Gunn and D. H. Weinberg, ``The Sloan Digital Sky Survey" in {\it Wide-Field Spectroscopy and the Distant Universe}, ed. S. J. Addox and A. Arag\'on-Salamanca (World Scientific, Singapore), pp. 3-14 (1995).
\item{[4]} M. Chaves and H. Morales, Mod. Phys. Lett. {\bf 13A}, 2021 (1998).
\item{[5]} A. Yu. Kamenschik, I. M. Khalatnikov and A. V. Toporensky, Phys. Lett. {\bf B357}, 36 (1995); 
D. Scialom, Helv. Phys. Acta {\bf 69}, 190 (1996); 
P. Jetzer and D. Scialom, Phys. Rev. {\bf D55}, 7440 (1997); 
D. Scialom and P. Jetzer, Phys. Rev. {\bf D51}, 5698 (1995); 
L. Amendola, I. M. Khalatnikov, M. Litterio and F. Occhionero, {\bf D49}, 1881 (1994).
\item{[6]} M. Chaves, Phys. Lett. {\bf B415}, 175 (1997).
\item{[7]} A. D. Linde, Phys. Lett. {\bf 129B}, 177 (1983); 
A. D. Linde, Phys. Lett. {\bf 162B}, 281 (1985); 
A. D. Linde, {\it Inflation and Quantum Cosmology}, Academic Press, San Diego, CA. 1990.
\item{[8]} P. H. Coleman and L. Pietronero, Phys. Rep. {\bf 213}, 311 (1992).
\item{[9]} J. Mather {\it et al.}, Astrophys. J. {\bf 354}, L37 (1990); 
G. Smoot {\it et al.}, Astrophys. J. {\bf 396}, L1 (1992); 
C. Bennett {\it et al.}, Astrophys. J. {\bf 396}, L7 (1992); 
E. Wright {\it et al.}, Astrophys. J. {\bf 396}, L11 (1992).
\end